\documentstyle[12pt,aasms]{article}
\begin{document}
\def\be{\begin{equation}}
\def\en{\end{equation}}
\def\bea{\begin{eqnarray}}
\def\ena{\end{eqnarray}}
\title{
Neutrino pair synchrotron radiation from relativistic electrons
in strong magnetic fields}
\author{A. Vidaurre$^{1}$, A. P\'erez$^{2}$, H. Sivak $^4$,
J. Bernab\'eu$^{2,3}$,
and
J. M$^{\underline{\mbox{a}}}$. Ib\'a\~{n}ez$^{2}$}
\affil{
$^{1}$ Departamento de F\'{\i}sica Aplicada,
Universidad Polit\'ecnica de Valencia, Spain}
\affil{
$^{2}$ Departamento de F\'{\i}sica Te\'orica, Universidad de Valencia}
\affil{
$^3$ IFIC, Centro Mixto Univ. Valencia-CSIC, 46100 Burjassot (Valencia), Spain}
\affil{
$^4$ D.A.R.C., Observatoire de Paris-Meudon, 92190 Meudon, France}

\begin{abstract}

\noindent

The emissivity for the neutrino pair synchrotron radiation in strong magnetic
fields has been calculated both analytically and numerically for high densities
and moderate temperatures, as can be found in neutron stars. Under these
conditions, the electrons are relativistic and degenerate. We give here our
results in terms of an universal function of a single variable. For two
different regimes of the electron gas we present a simplified calculation and
compare our results to those of Kaminker et al. Agreement is found for the
classical region, where many Landau levels contribute to the emissivity , but
some differences arise in the quantum regime. One finds that the emissivity for
neutrino pair synchrotron radiation is competitive, and can be dominant, with
other neutrino processes for magnetic fields of the order $B \sim 10^{14} -
10^{15} G$. This indicates the relevance of this process for some astrophysical
scenarios, such as neutron stars and supernovae.

\vskip 0.25cm

\noindent
\it{Subject headings :} Stars : Magnetic Fields---Stars: Neutron.
\end{abstract}
\vskip 0.5cm

\section{INTRODUCTION}

Estimates of the magnetic field strength at the surface of neutron stars are
obtained from several different scenarios: theoretical models of pulsar
emission (\cite{Ru72}), the accretion flow in binary X-ray sources
(\cite{GL78}) and observation of features in the spectra of pulsating X-ray
sources which have been interpreted as cyclotron lines
(\cite{Tetal78,Wetal79,Getal80,Metal90}). For a recent review of all these
topics the interested reader is addressed to the book by Michel (\cite{Mi91}).
In a sample of more than 300 pulsars the range of values of the surface
magnetic field strength runs into the interval: $10.36\leq log B$ (Gauss) $\leq
13.33$ (\cite{MT81}).

Very recently, several authors (\cite{DT92,TD93,BKM92,BK93})  have
proposed two different physical mechanisms leading to an amplification of some
initial magnetic field in a collapsing star. Fields as strong as $B \sim
10^{14}-10^{16} G$, or even more, might be generated in new-born neutron stars.

According to (\cite{BKM92,BK93}), a mirror-asymmetric magnetic field
distribution might arise in a rapidly  and differentially rotating
proto-neutron star having, originally, both a toroidal and a poloidal
component. The field amplification due to differential rotation leads to the
formation of an additional toroidal field from the poloidal one by twisting of
the field lines. After the first 20 seconds of the life of a new-born neutron
star (basically, its Kelvin-Helmholtz epoch) the induced toroidal magnetic
field could be as huge as $B \sim 10^{15}-10^{17} G$.

In a second scenario (\cite{DT92,TD93}), a dynamo action in a differentially
rotating and convective young neutron star is responsible for the
strengthening of some initial dipole field up to values of $B \sim 10^{12}-3
\times 10^{13} G$ if the convective episodes arose during the main-sequence
stage or to $B \sim 10^{14}-10^{15} G$ if the dipole field is generated after
collapse.

In presence of a strong magnetic field
the so-called neutrino-pair synchrotron radiation process becomes allowed :
\begin{equation}
[e^{-}]\,\stackrel{\vec{B}}{\rightarrow}\,[e^{-}]\,+\,\nu\,+\,\bar{\nu}
\label{syn}
\end{equation}

This reaction has been studied by a few groups
(\cite{L67,Ca70,Ya81,Vi90,Ka91,Ka92,Ka93}) for different regimes of the
electron
plasma. The calculation of the corresponding emissivity, both analytically or
numerically, is far from obvious. In fact, this calculation appears as a
multiple integral and summation over the variables and quantum numbers
involving the wave functions of the initial and final electron and the
corresponding statistical weights. The wave function integrals lead to Laguerre
or Bessel functions with a complicated behaviour, so that approximations in
order to simplify the expressions are sometimes delicate. This in fact has lead
to errors in the past literature.

In a recent paper, Kaminker et al. (\cite{Ka91}) have studied the
neutrino emissivity of the above process (\ref{syn}) for moderately high
magnetic fields $B \sim 10^{12}-10^{14} G$ and high densities, in the
degenerate-relativistic regime for the electrons. They claim that, within these
conditions, the emissivity is independent of the electron density. So far,
to our knowledge, there are no numerical tests which confirm these results.

Given all these circumstances, together with the interest of the problem, both
from the theoretical and astrophysical point of view, we have considered useful
to reexamine the existing results for the above regime. Therefore, we have
performed a numerical study of this process at high densities $\rho Y_e \ge
10^7 g/cm^3$, where $Y_e$ is the electron fraction per baryon and $\rho$ the
matter density in c.g.s. units, and moderate temperatures $T < 10^9 K$, for
values of the magnetic field strength $B \le 10^{16} G$. Under these
conditions, the electrons are relativistic and degenerate. We have found a
result which is in agreement with the one of Kaminker et al. when the electron
gas is in the classical regime. We also obtain agreement for the corresponding
analytical expressions, which we present in a simpler way than these authors.
For the quantum regime, however, we give analytical formulae which show the
correct dependence on $B$ for large magnetic fields, in contrast to those given
by Kaminker et al. This fact would be particularly important if some of the
magnetic field amplification mechanisms described above were physically
realizable in nature.

The influence of the neutrino-pair synchrotron radiation in presence of strong
magnetic fields merits to be examined in different astrophysical scenarios,
such as the delayed mechanism of type II Supernovae, neutrino cooling of
proto-neutron stars during the Kelvin epoch or the secular cooling of the
neutron star. Furthermore, the neutrinos originated from the combined effect of
one of the proposed mechanisms to enhance the magnetic field and the
neutrino-pair synchrotron radiation process could be envisaged as a signature
of the mechanism itself.

This paper is organized as follows. In section 2 we discuss the calculation of
the emissivity for the synchrotron process and present our main results. These
results are illustrated with more detail in section 3 and 4 for two different
regimes of the electron gas. We end in section 5 with some conclusions and
remarks.

\section{CALCULATION OF THE EMISSIVITY}

The emissivity for the process (\ref{syn}) can be written as

\bea
\varepsilon_{\nu}\,= & \,\frac{G^{2}eB}{3(2\pi)^{6}}\,\,
\sum^{\infty}_{n=1} \sum^{n-1}_{n'=0}\,\,\int^{+\infty}_{-\infty}dp_z\,
\int_{\vec q^{2} \leq \omega^2 } d^{3}q\ \omega \nonumber \\
& \times\,\, A\,f(E)\,[1\,-\,f(E')]
\label{emis1}
\ena

\noindent
where $G$ is the Fermi coupling constant. The initial (final) electron, of mass
$m$, has an energy $E=\sqrt{m^2 + p_\perp^2 + p_z^2}$ ($E'=\sqrt{m^2 +
p'^2_\perp + p'^2_z}$) characterized by the Landau quantum number $n$ ($n'$)
and momentum $p_z$ ($p'_z$) along the B-direction. We have introduced
$p_\perp^2=2 e B n$ and $p'^2_\perp = 2 e B n'$, which correspond to the
classical transverse momenta of the electron. In the above equation, $q$ is the
four-momentum transfer and $\omega = E - E'$ the energy which is carried away
by the neutrino pair. $f(E)$ is the Fermi-Dirac distribution function : $f(E)\
=\ \left[ \exp (\frac{E-\mu}{T}) + 1 \right] ^{-1}$, where $\mu$ is the
electron chemical potential. The integration region over $\vec{q}$ is
restricted by :

\be
q^2\,=\,\omega^{2}\,-\,\vec q ^2\,\geq \,0.
\label{phase}
\en

The expression for $A$ in the equation of the emissivity can be found in
(\cite{Vi90,Ka92}).
In the relativistic limit, the relevant expression for $A$ is :

\bea
A & = & \frac{(C_V^2 + C_A^2)}{E E'} \left\{ [ -2(EE'-p_z p'_z)^2
\right.  \nonumber \\
& + & (EE'-p_z p'_z) (p_t^2 - 2 q_\perp ^2)
- \left. \frac{q_\perp ^2}{2} (q_\perp^2 - p_t^2) \right] \Psi (u) \nonumber
\\
& + & p_t^2 \left[ (EE'-p_z p'_z) \right. -  \left. \left.  \frac{1}{2}
(p_t^2 - q_\perp ^2) \right] \Phi (u) \right\}
\label{A}
\ena

\noindent where

\bea
\Psi & = & \frac{n'!}{n!} u^s e^{-u}
[\frac{u}{n'} (L^{s+1}_{n'-1})^2 + \frac{n}{u} (L^{s-1}_{n'})^2]  \nonumber \\
\Phi & = & \frac{n'!}{n!} u^s e^{-u}
[\frac{n}{n'} (L^s_{n'-1})^2 + (L^s_{n'})^2] \label{lag}
\ena

$L^{n_2}_{n_1}$ are Laguerre functions with argument $u=\frac{q_\perp^2}{2eB}$
($q_\perp$ is the component of $\vec q$ orthogonal to the magnetic field) and
$s=n-n'$. $C_V$ ($C_A$) is the effective vector (axial) coupling of the
neutrino pair to the electron current, coming from both the Fierz reordered
charged current (for electron neutrinos) and neutral current (for all neutrino
species) weak interactions. In Eq. (\ref{A}) we have dropped a term
proportional to $(C_V^2 - C_A^2)$, which disappears in the extreme relativistic
limit, and the interference term proportional to
$C_V C_A$, which does not contribute to the integrated emissivity in Eq.
(\ref{emis1}). The argument goes as follows. The corresponding integrand,
for $C_V C_A$, is odd under the simultaneous change of sign of the longitudinal
momenta of the initial and final electrons, so that a symmetric integration
of both $p_z$ and $q_z$ in Eq.(\ref{emis1}) cancels this asymmetric term.
We have defined $p_t^2 =
p_\perp ^2 + p'^2_\perp$. If one takes for the electroweak mixing angle
$ sin^2 \theta_W = 0.23 $, then $ C_V^2 + C_A^2 = 1.6748$.

One has the relationship


\be
\omega^2 - q_z^2 = 2m^2 + p_t^2 -2(EE'-p_z p'_z) \approx p_t^2 -2(EE'-p_z p'_z)
\en


where the latter approximation corresponds to considering relativistic
electrons.

By substituting in Eq. (\ref{A}) we obtain


\be
A  =  \frac{(C_V^2 + C_A^2)}{2 E E'} (\omega^2-q_z^2-q^2_{\perp})[p_t^2
(\Psi-\Phi) - (\omega^2-q_z^2-q^2_{\perp}) \Psi ]
\label{A2}
\en


As mentioned above, for the range of temperatures and densities we are
interested in, the electrons are degenerate. The product of
distribution functions appearing
in (\ref{emis1}) will then restrict the energies to $E,E' \sim \mu$.
 Moreover, one can
write the following identity :

\be
f(E)\,[1\,-\,f(E')] = {\cal B}(\omega) \left[ f(E') - f(E) \right]
\label{bose}
\en

\noindent where ${\cal B}(\omega) = [ \exp(\omega/T) - 1 ]^{-1}$ is a
Bose-Einstein distribution function with zero chemical potential. From the last
equation, it is apparent that the energy difference $\omega$ will be restricted
to a few times $T$ and will be much lower than the relevant values of $E$ and
$E'$, if the electrons are degenerate. Similarly, one can easily estimate
that $q_z$ and $q_{\perp}$ will contribute as $\sim T$, whereas $p_t$, $p_z$
and
$p'_z$ contribute as $\sim \mu$.
This allows us to neglect the last term in Eq. (\ref{A2}). In this case one
gets

\be
A  =  \frac{(C_V^2 + C_A^2)}{2 E E'} p_t^2 (\omega^2-q_z^2-q^2_{\perp}) \Theta
\label{A3}
\en

with $\Theta=\Psi-\Phi$. We have tested numerically that
the complete expression Eq.(\ref{A}) or Eq.(\ref{A2}) gives
approximately the same result as Eq.(\ref{A3}) for the physical conditions
we are considering here.

We can also perform the following approximations, in agreement with the
above discussion :


\bea
n+n'   & = & \frac{p_t^2}{2 e B} = \frac{(E+E')^2 + \omega^2
           -2(p_z^2+p'^2_z)}{4eB} \approx
           \frac{(E+E')^2  -4 p_z^2}{4eB} \nonumber \\
\omega & = & \frac{2 s e B + p_z^2 -p'^2_z}{E+E'} \approx
             \frac{2 s e B + 2 p_z q_z}{E+E'}
\label{apro1}
\ena


Due to the Pauli principle, the quantity $f(E)
[1-f(E')]$ will be nonzero only when $E$ and $E'$ are within a narrow interval
around $\mu$ of width $\omega \sim T$. With this in mind, we have replaced all
the slowly varying functions in (\ref{emis1}) by their value around $E, E' \sim
\mu$.

The latter equation, together with the phase space restriction (\ref{phase}),
requires that $q_z$ and $q_{\perp}$ must lie inside the elliptical domain
(\cite{Ka91})

\be
q^2_{\perp} + \frac{p^2_{\perp}}{\mu^2} (q_z - \chi_z)^2 \le \chi^2_{\perp}
\label{ellipse}
\en

\noindent where $\chi_z = \frac{s e B p_z}{p^2_{\perp}}$ and $\chi_{\perp}
= \frac{s e B}{p_{\perp}}$.

We have numerically calculated the emissivity of the synchrotron process
(\ref{syn}) for values of the electron density $\rho Y_e \ge 10^7 g/cm^3$,
moderate temperatures $T < 10^9 K$, and magnetic fields $B \le 10^{16} G$. The
results, for this range of values, can be expressed in a compact way as :

\be
\varepsilon_{\nu}\,= 1.47\ 10^{14}\ B^2_{13}\ T^5_9\ f(x)\  erg/cm^3/s
\label{f}
\en

\noindent where $B_{13}$ is the magnetic field in units of $10^{13} G$, $T_9$
is the temperature in units of $10^9 K$, and $x\ =\ \frac{\mu T}{e B}$ is a
dimensionless variable. This result is useful in order to perform analytical
calculations. We have plotted in Fig. 1 the function $f(x)$. As can be seen
from this figure, $f(x)$ first increases as $x$ grows, and is almost constant
for large values of $x$. This behaviour corresponds to different physical
regimes of the electron plasma, and will be explained in the next sections.

\section{Large x Regime}

Let us discuss the behaviour of the emissivity corresponding to large values of
the parameter $x$ defined above. This situation  corresponds, for example, to
sufficiently high electron densities for a fixed magnetic field and
temperature.
The number of Landau levels which are involved in Eq. (\ref{emis1}) is $n_{max}
= \frac{\mu^2}{2 e B}$. On the other hand, the maximum of $s = n -n'$ can be
estimated from Eq. (\ref{apro1}) as $s_{max} \sim \frac{\mu T}{eB} =x$. We then
have  $n_{max} >> s_{max} >> 1$ for $\frac{\mu}{T} >> 1$. According to this
idea, we have used the following approximations for the Laguerre polynomials in
Eq. (\ref{lag}) :

\bea
\Psi & \rightarrow & J_{s-1}^2 + J_{s+1}^2 \nonumber \\
\Phi & \rightarrow & 2 J_s^2
\label{bes}
\ena

\noindent
with $J_s(a)$ a Bessel function and $a = \sqrt{2(n+n') u}$. One can prove that
$a \le s$ always. The approximation shown by Eq. (\ref{bes}) can be used if $n
>> 1 $ and $s << n$. Because it is time saving and more suitable numerically
than Eq. (\ref{lag}), we made use of it in our numerical computation of the
emissivity, whenever large values of $n$ (and $s << n$) were encountered. By
changing the sum over $n$ and $n'$ to a sum over $n'$ and $s$, and integrating
the angle of $\vec q$ around the $\vec B$ direction, one arrives to the
following expression :


\bea
\varepsilon_{\nu}\,= & \,\frac{G^{2}eB}{3(2\pi)^{5}}\, (C_V^2 + C_A^2)\,
\sum^{\infty}_{n'=0} \sum^{\infty}_{s=1}\,\,\int^{+\infty}_{-\infty}dp_z\,
\int_{q^{2} \leq \omega^2} dq_z dq_{\perp} q_{\perp} \ \omega \nonumber \\ &
\times\,\, (1-\frac{p^2_z}{\mu^2}) (\omega^2-q_z^2-q^2_{\perp})\ {\cal B}
(\omega) \left[ f(E') - f(E) \right] \Theta (a)
\label{emis2}
\ena


\noindent
The argument of $\Theta$ can be written as $a = p_{\perp}  q_{\perp}/(eB)$,
with
$p_{\perp}=\sqrt{\mu^2 - p_z^2}$, and $\omega = \frac{s e B + p_z q_z}{\mu}$.
Further approximations can be made in the above equation for degenerate
electrons, if one considers the distribution functions $f(E)$ and $f(E')$ as
step functions in the energy. Within this assumption, the sum over $n'$ can be
done explicitly. One gets

\be
\sum^{\infty}_{n'=0} \left[ f(E') - f(E) \right] =\ \frac{\omega}{2eB} (E+E')
\simeq \frac{\mu \omega}{eB}
\label{sumnp}
\en

In deriving Eq.(\ref{sumnp}), we have used the fact that $s$ is lower
than $n_{max}$. By inserting the latter
expression into Eq. (\ref{emis2}) we obtain


\bea
\varepsilon_{\nu}\,= & \,\frac{G^{2} \mu}{3(2\pi)^{5}}\, (C_V^2 + C_A^2)\,
\sum^{\infty}_{s=1}\,\,\int^{+\infty}_{-\infty}dp_z\,
\int_{q^{2} \leq \omega^2} dq_z dq_{\perp} q_{\perp} \ \omega^2 \nonumber \\ &
\times\,\, (1-\frac{p^2_z}{\mu^2}) (\omega^2-q_z^2-q^2_{\perp})\
{\cal B}(\omega) \Theta (a)
\label{emis3}
\ena


Next we make use of recurrence relationships for the Bessel functions and
obtain the formula :

\be
\Theta(a) = 2 (J'_s)^2 + 2 (s^2/a^2-1) J^2_s
\label{theta1}
\en

For the large values of $s$ involved here, one can use the following
approximation to the Bessel functions (\cite{GR80}):

\be
J_s(a) \approx \frac{1}{\pi} \sqrt{\frac{2(s-a)}{3a}} K_{1/3} (z)
\en

where $z=\frac{[2(s-a)]^{3/2}}{3 \sqrt{x}}$ and $K_{1/3}$ is the modified
Bessel function, which can be further approximated as :

\be
K_{1/3} (z) \approx \sqrt{\frac{\pi}{2z}} \exp{(-z)}
\en

In this way one obtains, after some algebra,

\be
\Theta(a) \approx \frac{2}{\pi} s^{-4/3} (2 z)^{1/3} \exp{(-2 z)}
\label{theta2}
\en

As can be seen from
Fig. 2, the latter equation provides a reasonable approximation to
Eq.(\ref{theta1}). In this figure,
 we have plotted $\Theta(a)$ as obtained from Eq.(\ref{theta2})
(dotted line) and from Eq.(\ref{theta1}) (solid line) for $s=200$. Another
important feature is that only values of the argument $a$ close to $s$
will contribute. This can be understood from Eq.(\ref{theta2}), due to the
exponential behaviour, which effectively limits $z$. In fact, if we
define $\sigma$ as the 'width' of the exponential, one has
$(1-a/s) < (\frac{\sigma}{s})^{2/3}$. By substituting into Eq.(\ref{ellipse})
one obtains that important values of $q_z$ and $q_{\perp}$ are restricted to

\bea
(1-\frac{q_{\perp}}{\chi_{\perp}}) & \le & (\frac{\sigma}{s})^{2/3} \nonumber
\\
|q_z - \chi_z| & \le & \Delta(q_{\perp})  \le \sqrt{2}\ \frac{\mu \chi_z}{p_z}\
(\frac{\sigma}{s})^{1/3}
\ena

We have defined $\Delta(q_{\perp}) = \frac{\mu \chi_{\perp}}{p_{\perp}}
\sqrt{1-\frac{q^2_{\perp}}{\chi^2_{\perp}}}$. Thus $q_z$ is restricted to a
narrow interval of width $\Delta(q_{\perp})$ around $\chi_z$. This allows us to
perform the integral over $q_z$ approximately. To the first order in
$\Delta(q_{\perp})$ one can write

\be
\int_{\chi_z - \Delta(q_{\perp})}^{\chi_z + \Delta(q_{\perp})} dq_z
\longrightarrow 2 \Delta(q_{\perp})
\en

with the replacement $q_z \rightarrow \chi_z$ in the integrand of
Eq. (\ref{emis3}). This means that $\omega^2$ will be replaced by
$\chi^2_z + \chi^2_{\perp}$. The integral over $q_{\perp}$ is then
immediate. Since the total number of level differences $s$ is large, one
can substitute the sum over $s$ by an integral over the continuous variable
$t = s/x$. Thus by changing $\sum_{s} \longrightarrow x \int^{\infty}_{0} dt$
and performing the remaining integrals one finally obtains


\be
\varepsilon_{\nu} = \frac{G^{2}}{\pi^{6}}\, (C_V^2 + C_A^2) \zeta(5)
(e B)^2 T^5 = 1.16\ 10^{15}\ B_{13}\ T^5_9\ erg/cm^3/s
\label{final}
\en


This equation implies that the emissivity does not depend on the electron
density in this regime, in agreement with the result previously found in Ref.
(\cite{Ka91}). In fact, our Eq. (\ref{final}) is close to the one obtained in
this reference. It is also in good agreement with the values of $f(x)$ obtained
numerically (and plotted in Fig. 1), as can be seen by comparing
Eq.(\ref{final}) with Eq.(\ref{f}).

A numerical fit which reproduces the behaviour of $f(x)$ for $x>2$ to better
than 4\% is given by
\be
        f(x)= \frac{-5.0224 - 8.1289 x + 9.2892 x^2}
                    {1.0293 + 2.0605 x + 1.0727 x^2}
\en

\section{Low x Regime}

We now address the question of whether an increase of the magnetic field will
always give a larger neutrino emission. In Fig. 3 we present the corresponding
emissivity (in cgs units and logarithmic scale) for a fixed temperature $T_9 =
1$ and four values of the electron density $\rho Y_e = 10^9, 10^{11}, 10^{13},
10^{14} g/cm^3$, as a function of $B_{13}$ ($B_{13}$ ranging from unity up to
$10^3)$, as obtained numerically. For these high magnetic fields, the number of
populated Landau levels ($n_{max}$, defined above) can be of order unity, and
we enter into the quantum regime. We have used, in these cases, the expression
of $A$ as given by Eqs. (\ref{A}) and (\ref{lag}) directly, instead of making
the approximations shown in section 3.

\noindent
As can be seen from Fig. 3, for a given density  and temperature, the
emissivity first increases
and, after reaching a maximum value, will fall to zero for large values of the
magnetic field. This can be understood since the number of
possible $n \rightarrow n'$ transitions decreases as $\frac{eB}{\mu T} \gg 1$.
In fact, a rough analytic expression in this region can be obtained by putting
$s=1$ in Eq. (\ref{emis2}) and $\omega = \frac{eB}{\mu}$. Therefore one has :

\bea
\Psi & \simeq & 1 \nonumber \\
\Phi & \simeq & 0
\ena

The integrals in Eq. (\ref{emis2}) can be done analytically and one obtains the
following expression :

\be
\varepsilon_{\nu}\,= 6.6\ 10^{12}\ B^2_{13}\ T^5_9\ x^{-5} e^{-1/x}\
erg/cm^3/s
\label{expo}
\en

We have verified that the latter expression gives values which are in agreement
with our numerical results around the
maximum of the emissivity. This is shown in Fig. 4, where we have compared our
results for $\rho Y_e = 10^{11} g/cm^3$ (solid line) with the
prediction of Eq. (\ref{expo}) (dotted line). We have also plotted (dashed
line)
the corresponding analytical approximation given by (\cite{Ka91}) for this
case.
As can be seen from this figure, Eq. (\ref{expo}) provides a reasonable
approximation to the emissivity, which works better than the formula of
Kaminker
et al.

\section{Comparison with other processes}

In order to investigate the relevance of the process studied here for
neutron star cooling, we
have made the comparison with the emissivities corresponding to other neutrino
processes which are competitive with the synchrotron emission. We
have considered pair production $e^+ e^- \longrightarrow \nu \bar{\nu}$
,plasmon decay $\Gamma \longrightarrow \nu \bar{\nu}$,
bremmstrahlung $e^- (Z,A) \longrightarrow e^- (Z,A) \nu \bar{\nu}$
and photoproduction $\gamma e^- \longrightarrow e^- \nu \bar{\nu}$.
For these processes, we have made
the assumption that they do not vary significantly with the magnetic field.
The
numerical fit to these emissivities have been taken from \cite{mu85}.
Although these rather simple formulae are not the most up-to-date available
fits to the above processes, they serve to our purpose as a first
approximation (see, for example,  \cite{mu89}, for a more
elaborated fit). The bremmstrahlung process is taken from  \cite{ma79}.
A more complete calculation, as in Itoh et al.
(\cite{mu89}), gives the same order of magnitude
in the region where this process dominates.

In Fig. 5 we present the result of comparing all these processes for a
temperature $T = 10^8 K$ as the product $\rho Y_e$ varies from zero up to
$10^{12} g/cm^3$. The bremmstrahlung energy emission was calculated
assuming that the dominant nucleus is $^{56}Fe$. The synchrotron emissivity
is plotted (solid lines) for two values of the magnetic field : $B_{13}=10$
and $B_{13}=100$. Other relevant processes are : bremmstrahlung (dashed-dotted
line), plasma (long dash), and photoneutrino emission (short dash).
As can be seen from this figure, the synchrotron emission is competitive with
the above processes within the range $\rho Y_e \sim 10^9 - 10^{12}$. Moreover,
as pointed out by Pethick and Thorsson (\cite{pe93}), band-structure effects
can
suppress bremmstrahlung by a factor of 10 or more for temperatures less than
about $10^9 K$. In this case, the synchrotron emission would be the dominant
process in the above electron density range, if the magnetic field reaches
values of the order $\sim 10^{14} G$.

For higher temperatures, the synchrotron emission
corresponding to a given value of $B$ 'switches on' at lower
electron densities, as can be inferred from Eq. (\ref{expo}). However,
other processes have a faster increase with temperature and, therefore,
the dominance of synchrotron emission reduces to a narrow interval of
densities,
although it effectively competes for high densities.
This is shown in Fig. 6, where we have made the above comparison for
a temperature $T = 10^9 K$ (pair emission is represented by the dotted line).

\section{CONCLUSIONS}

We have performed numerical calculations of the synchrotron emissivity from
relativistic degenerate electrons. This calculations allow us to present the
results in terms of an universal function $f(x)$ which can be used in
astrophysical codes. For two different regimes of the electron gas we have
derived analytical formulae, in a simpler way than previous references. These
formulae have been tested, and we have found a reasonable agreement with our
numerical calculations. We also have compared our results to the analytical
formulae derived recently by Kaminker et al. Agreement is found for the
classical region, where many Landau levels contribute to the emissivity , but
some differences arise in the quantum regime, for the values of the magnetic
field recently suggested in new-born neutron stars.

We have shown that neutrino-pair synchrotron radiation for moderately high
magnetic fields $B \geq 10^{14} G$ is an efficient cooling mechanism for
temperatures not larger than $10^9 K$, and can compete effectively (or even
dominate) with other processes.

We claim that the influence of the
neutrino-pair synchrotron radiation in presence of strong magnetic fields ($
\sim 10^{15} G$) merits to be examined in different astrophysical scenarios,
such as the delayed mechanism of type II Supernovae, neutrino cooling of
proto-neutron stars during the Kelvin epoch or the secular cooling of the
neutron star. Furthermore, these neutrinos originated from the combined effect
of the dynamo action with the neutrino-pair synchrotron radiation process could
be envisaged as a signature of this mechanism. This will be the subject of
future investigations.

{\bf Acknowledgments}
This work has been partially supported by the  Spanish DGICYT (grant
PB91-0648) and CICYT (grant AEN 93-0234).
Calculations were carried out in a VAX 6000/410 at the Instituto de
F\'{\i}sica Corpuscular and in a IBM 30-9021 VF at the Centre de
Inform\`atica de la Universitat de Val\`encia. We are grateful to J.A. Miralles
for useful comments.


\newpage

{\bf Figure Captions}

{\bf Figure 1.-}
The function $f(x)$ appearing in Eq. (\ref{f}). See the text for the definition
of the dimensionless variable $x$ and comments about its behaviour.
\vskip 0.5cm
\noindent

{\bf Figure 2.-}
The function $\Theta(a)$ as obtained from the approximation Eq.(\ref{theta2})
(dotted line) compared to Eq.(\ref{theta1}) (solid line) for $s=200$.

{\bf Figure 3.-}
Neutrino synchrotron emissivity as a function of the magnetic field $B_{13}$
for different values of the electron density. The temperature is the same
($T = 10^9 K$) in all cases.

{\bf Figure 4.-}
Comparison of our numerical
results for $\rho Y_e = 10^{11} g/cm^3$ (solid line) with the analytical
approximation Eq. (\ref{expo}) (dotted line). We have also plotted (dashed
line)
the approximation given by (\cite{Ka91}) for this case.

{\bf Figure 5.-}
Competition of synchrotron neutrino emission (solid lines) with other
processes, as a function of the electron density, for a temperature $T=10^8 K$.
Two different values of the magnetic field ($B_{13}=10$ and $B_{13}=100$
have been considered.
The bremmstrahlung emissivity (dash-dotted line) has been calculated for
$^{56}Fe$. Plasma process
is represented by long-dashed line, and photoneutrino by short-dashed line.

{\bf Figure 6.-}
Same as figure 5 for $T=10^9 K$. Dotted line corresponds to pair emission.

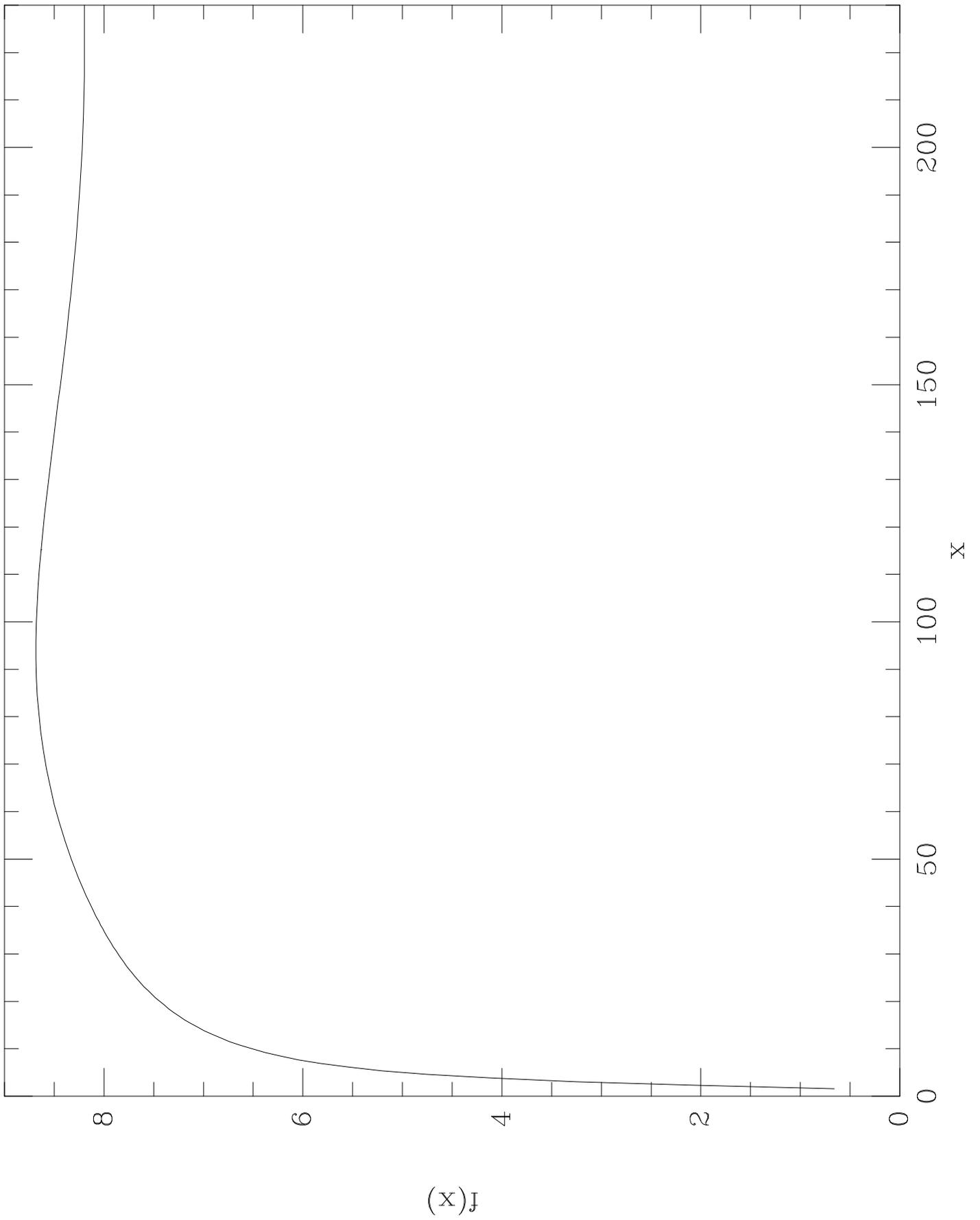
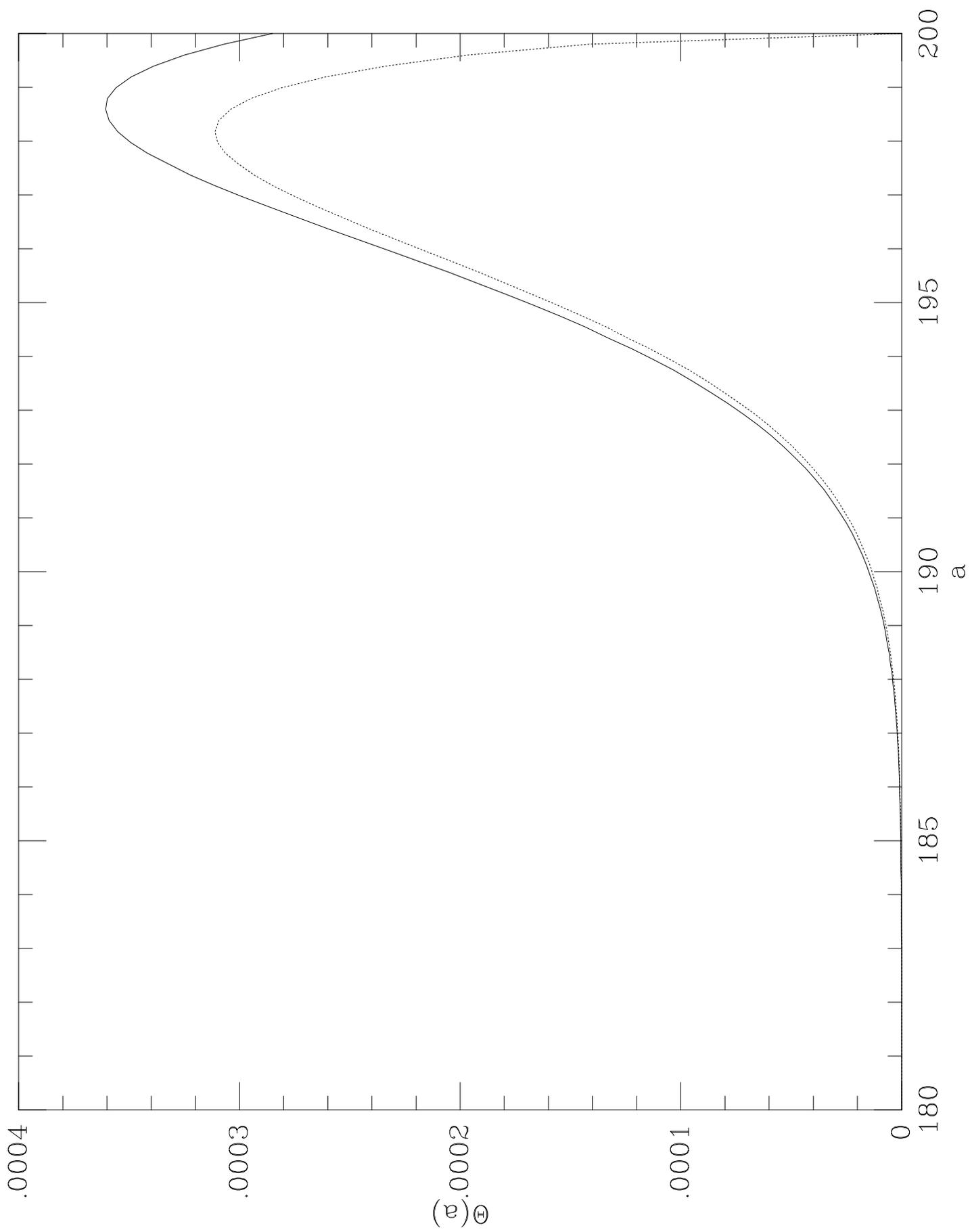
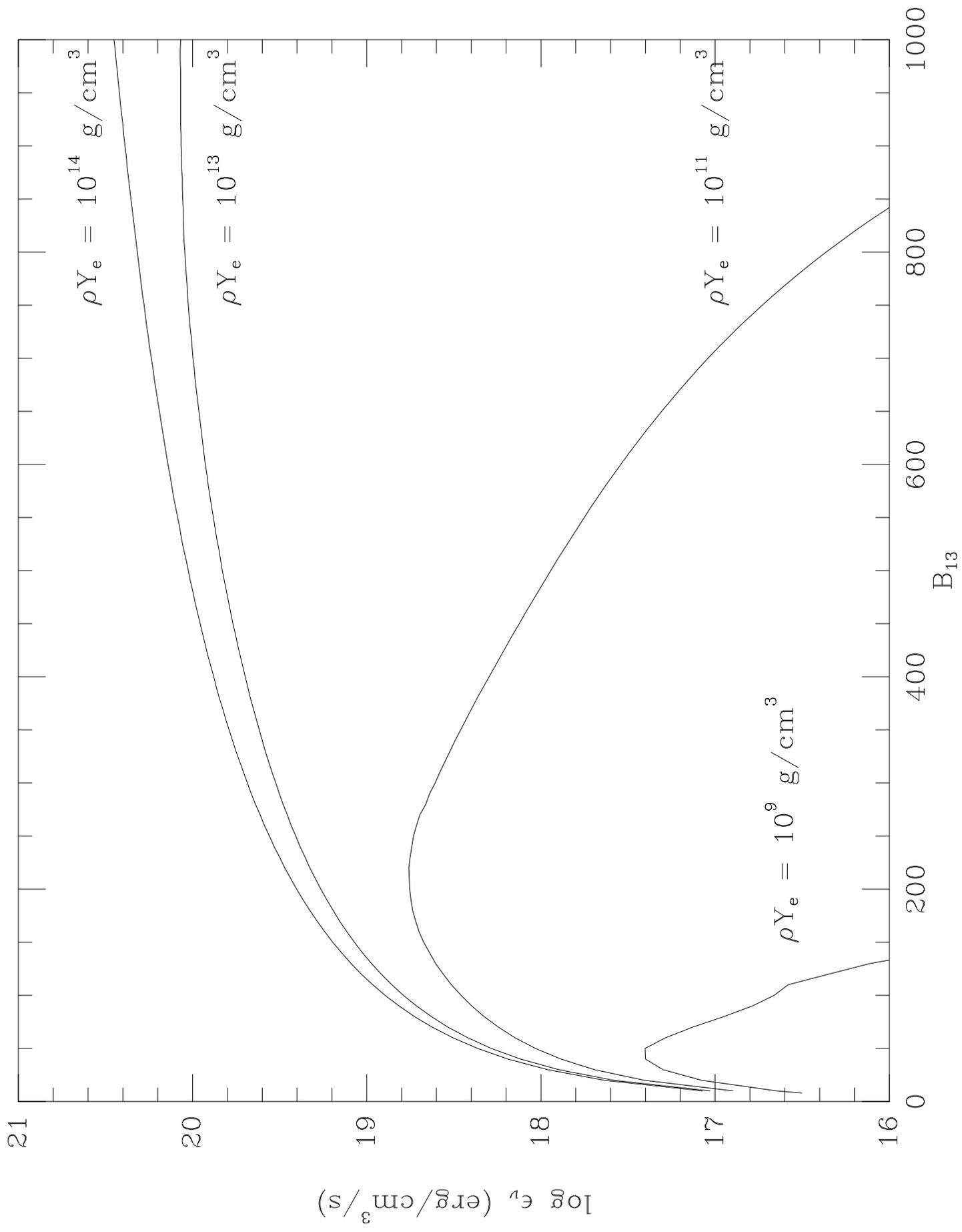
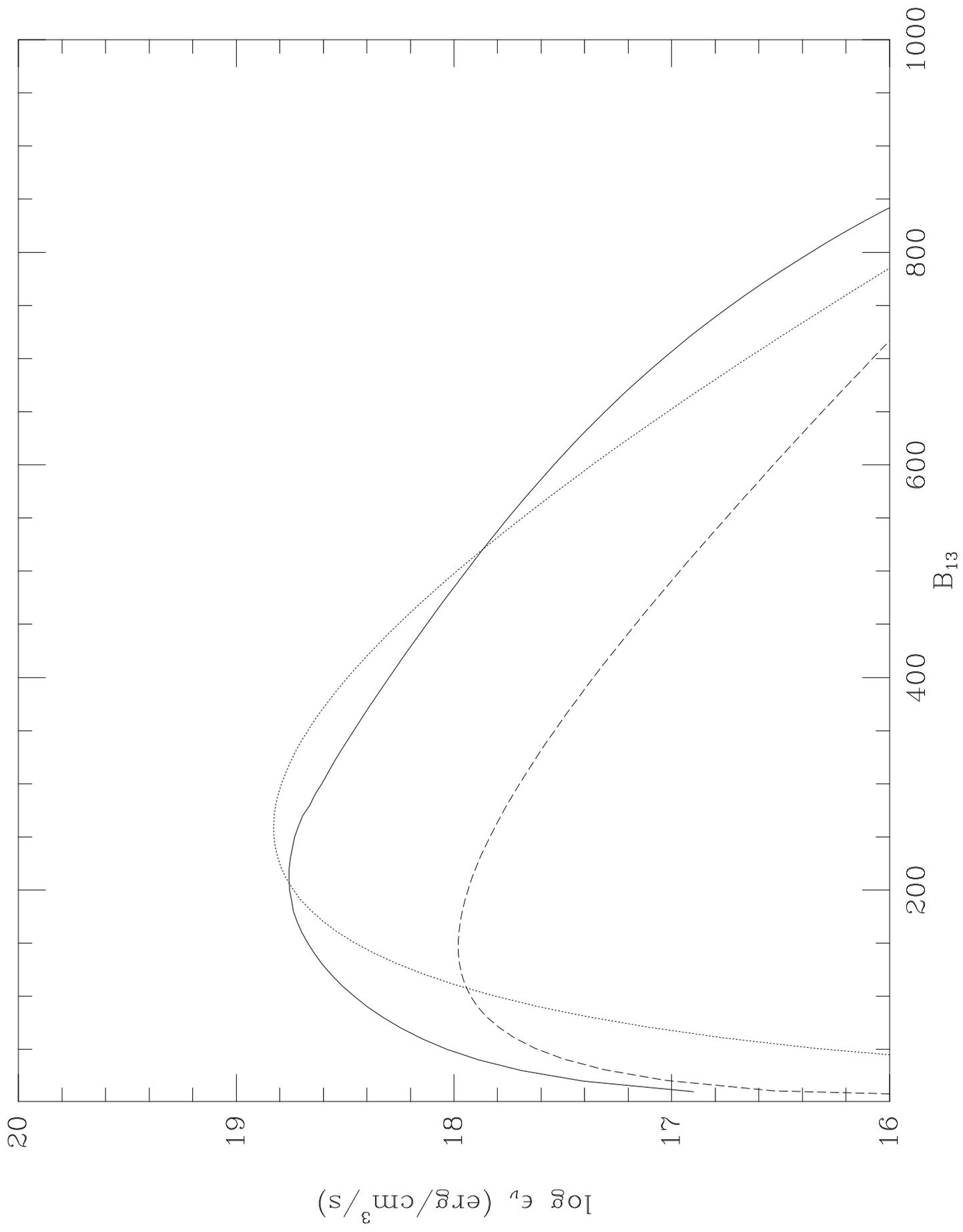
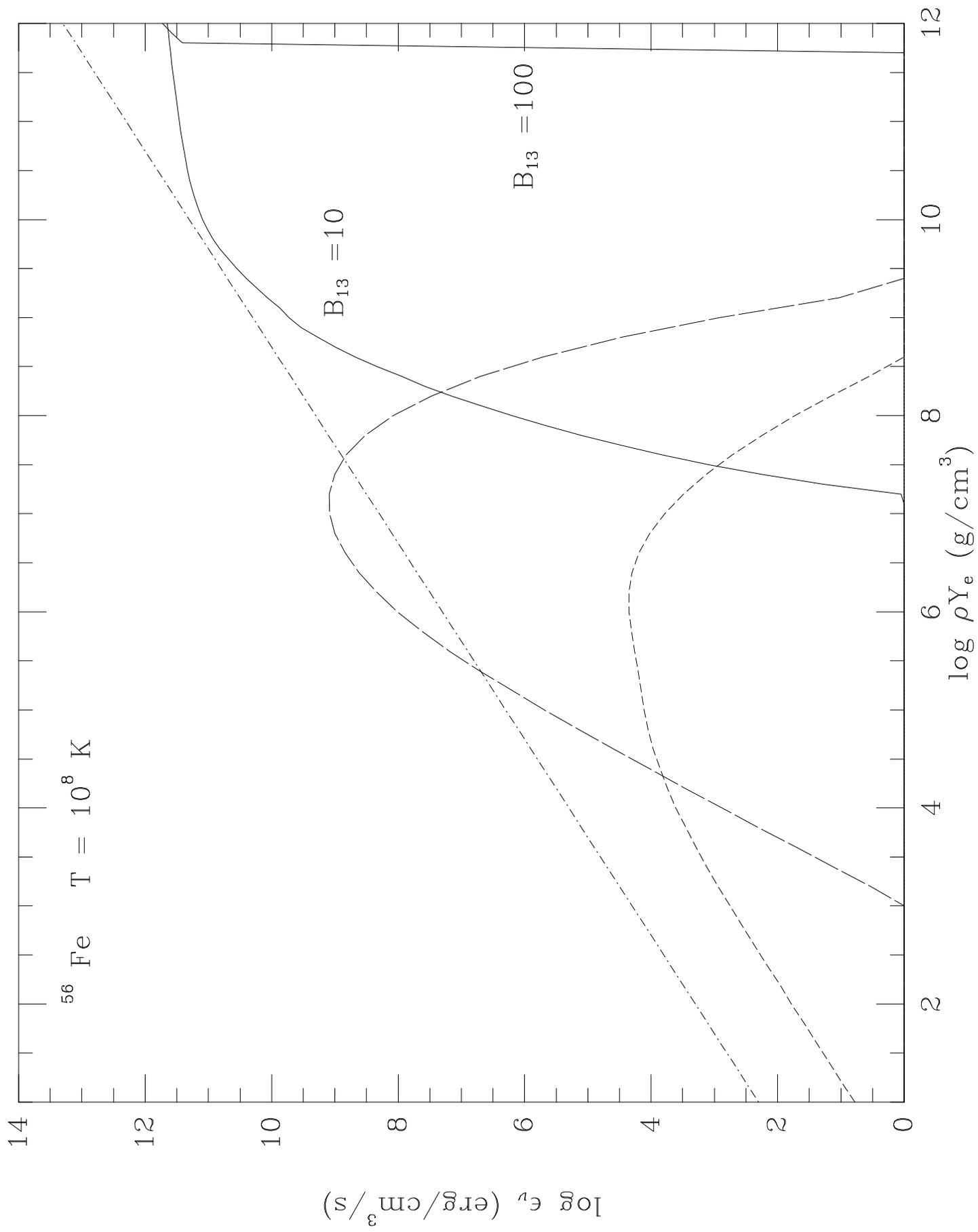
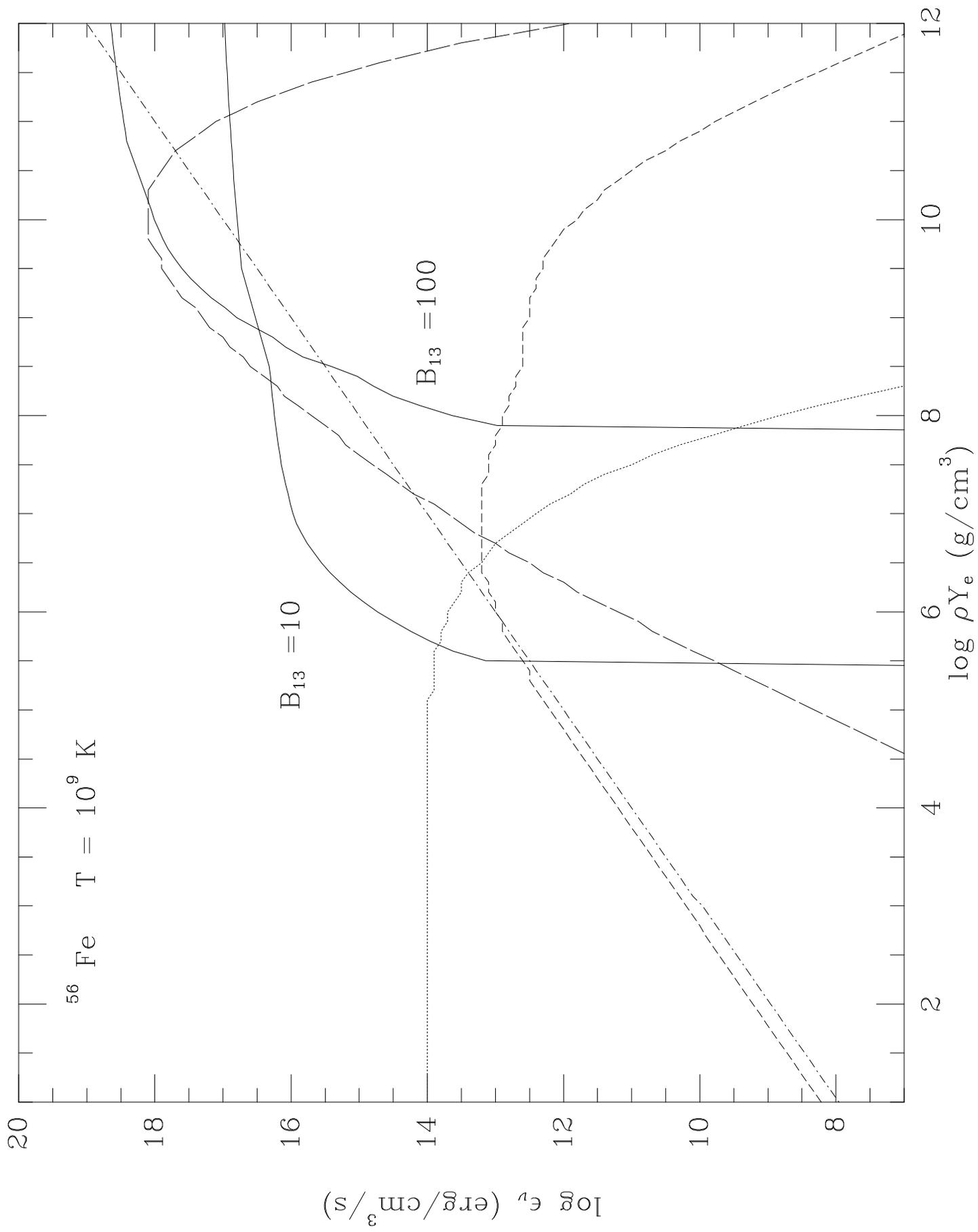
\end{document}